\documentclass[nocompress]{spie}  

\usepackage{amsmath,amsfonts,amssymb}
\usepackage{graphicx}
\usepackage[colorlinks=true, allcolors=blue]{hyperref}
\title{Software infrastructure for the highly-distributed semi-autonomous Dragonfly Spectral Line Mapper}

\author[a]{Imad Pasha}
\author[b,d]{Seery Chen}
\author[c]{Deborah Lokhorst}
\author[a]{William P. Bowman}
\author[a]{Zili Shen}
\author[b,d]{Qing Liu}
\author[e]{Evgeni I. Malakhov}
\author[b,d]{Roberto Abraham}
\author[a]{Pieter van Dokkum}

\affil[a]{Department of Astronomy, Yale University, 219 Prospect Street, New Haven, CT 06511}
\affil[b]{Dunlap Institute, University of Toronto, 50 St. George Street, Toronto, ON M5S3H4, Canada}
\affil[c]{NRC Herzberg Astronomy \& Astrophysics Research Centre, 5071 West Saanich Road, Victoria, BC V9E2E7, Canada}
\affil[d]{David A. Dunlap Department of Astronomy \& Astrophysics, University of Toronto, 50 St.
George Street, Toronto, ON M5S3H4, Canada}
\affil[e]{New Mexico Skies, Inc., 9 Contentment Crest, Mayhill, NM 88339, USA}
\authorinfo{Further author information: (Send correspondence to Imad Pasha.)\\Imad Pasha.: E-mail: imad.pasha@yale.edu, Telephone: 1 203 432 3000}

\begin{document} 
\maketitle
\begin{abstract}
The Dragonfly Spectral Line Mapper (DSLM)  is a semi-autonomous, distributed-aperture based telescope design, featuring a modular setup of 120 Canon telephoto lenses, and equal numbers of ultra-narrowband filters, detectors, and other peripherals. Here we introduce the observatory software stack for this highly-distributed system. Its core is the Dragonfly Communication Protocol (DCP), a pure-Python hardware communication framework for standardized hardware interaction. On top of this are 120 REST-ful FastAPI web servers, hosted on Raspberry Pis attached to each unit, orchestrating command translation to the hardware and providing diagnostic feedback to a central control system running the global instrument control software. We discuss key features of this software suite, including docker containerization for environment management, class composition as a flexible framework for array commands, and a state machine algorithm which controls the telescope during autonomous observations.

\end{abstract}


\section{INTRODUCTION}
\label{sec:intro}  

Distributed-aperture telescopes (also called `arrays' or `mosaic' systems) are seeing increasing adoption across several astronomical sub-disciplines, providing benefits of modular growth in effective area, smaller footprints, and in some cases, optical benefits for certain astronomical applications. The use of a ``multi-telescope-array'' was pioneered by the Dragonfly Telephoto Array \cite{Abraham:2014}, which comprised two mounts hosting 24 Canon telephoto lenses each. Equipped with SDSS-$g$ and SDSS-$r$ filters, the Dragonfly Telephoto Array has since observed much of the SDSS survey footprint in the largely unexplored regime of ultra low surface brightness observations over large fields of view. The core science of Dragonfly was unlocked by a combination of the properties of the refracting telescopes (commercial telephoto lenses) with highly advanced nano-coatings to prevent scattering and internal reflections, and the multiplexing itself, which produces an extremely low effective focal ratio, producing numerous new results including the discovery of dark matter deficient galaxies \cite{vanDokkum:2018}. Since the construction and commissioning of Dragonfly, other array based telescopes have been built or proposed, e.g., the Condor Array, made up of six 180 mm-diameter refracting telescopes \cite{Lanzetta2023a}, and the Argus Array \cite{Law:2022}, which will, when built, host 900 telescopes. 

Beyond the scientific benefits that can be afforded by specific implementations of multiplexed designs, there are economic and logistical strengths as well. Telescope and instrument costs scale strongly with size (e.g., mirror diameter). The 48 lenses of the Dragonfly Telephoto Array are an effective 1-m telescope in terms of light gathering power, but can take advantage of numerous cost savings by leveraging off-the-shelf consumer electronics and parts. The modularity of such designs thus makes the cost of nearly-arbitrary expansion a linear, rather than quadratic or exponential prospect.

The benefits of distributed aperture telescopic arrays also come with a variety of challenges, the primary of which is the orchestration and management of many individual (and repeated) components at once. In this work, we present the software stack for the Dragonfly Spectral Line Mapper (DSLM), an instrument built on the Dragonfly design framework but which has 120 lenses and numerous additional peripheral components \cite{chen2022}. Whereas on the original Dragonfly, each individual lens-camera-focuser ``unit'' is controlled by an Intel compute stick with the software needing only to communicate with a camera and a focusing unit (which controls the lens), the DSLM features a camera, a Starchaser focusing device which also serves as a camera using a pickoff mirror, the lens itself via an Arduino, and the motor for a filter tilter, which controls the tilt of ultra-narrowband filters mounted at the front of each unit. In total, the instrument comprises around 700 active hardware components at a time, all of which must be in active communication with a controlling software. 

There are several technologies which can considerably aid in the orchestration of many similar (or identical) units of this type, the primary of which is the Docker framework \cite{merkel2014docker}. Docker is a containerization scheme which aims to solve two primary tasks related (generally) to the hosting of software: (1) the creation of isolated runtime environments, and (2) the scaling of individual applications. The second use case --- in which, e.g., a web application built within a Docker container could be instantiated on arbitrarily more servers or nodes to service more incoming requests --- is not of great benefit to distributed aperture telescopes. But the former use case, of creating isolated and identical runtimes for an arbitrary number of systems is highly beneficial for ensuring that everything from the high level code, to its dependencies, to the operating system and core libraries are all the same across the distributed system. There is an additional benefit to the use of the Docker framework within this setting: container orchestration. Due to its popularity and use in industry, numerous other tools have been built in an ecosystem surrounding containers, and Docker directly has tools which aid in maintaining or changing the state of many containers at once. This eases management cases such as starting, stopping, and restarting the software servers (as well as monitoring them). 

By providing each unit on a distributed aperture array a computer (in the case of DSLM, a lightweight Raspberry Pi), we are able to instantiate docker containers containing the full software stack for the instrument, and establish the connections to the hardware associated with each unit.

\section{THE DRAGONFLY SPECTRAL LINE MAPPER}

\begin{figure}
    \centering
    \includegraphics[width=\linewidth]{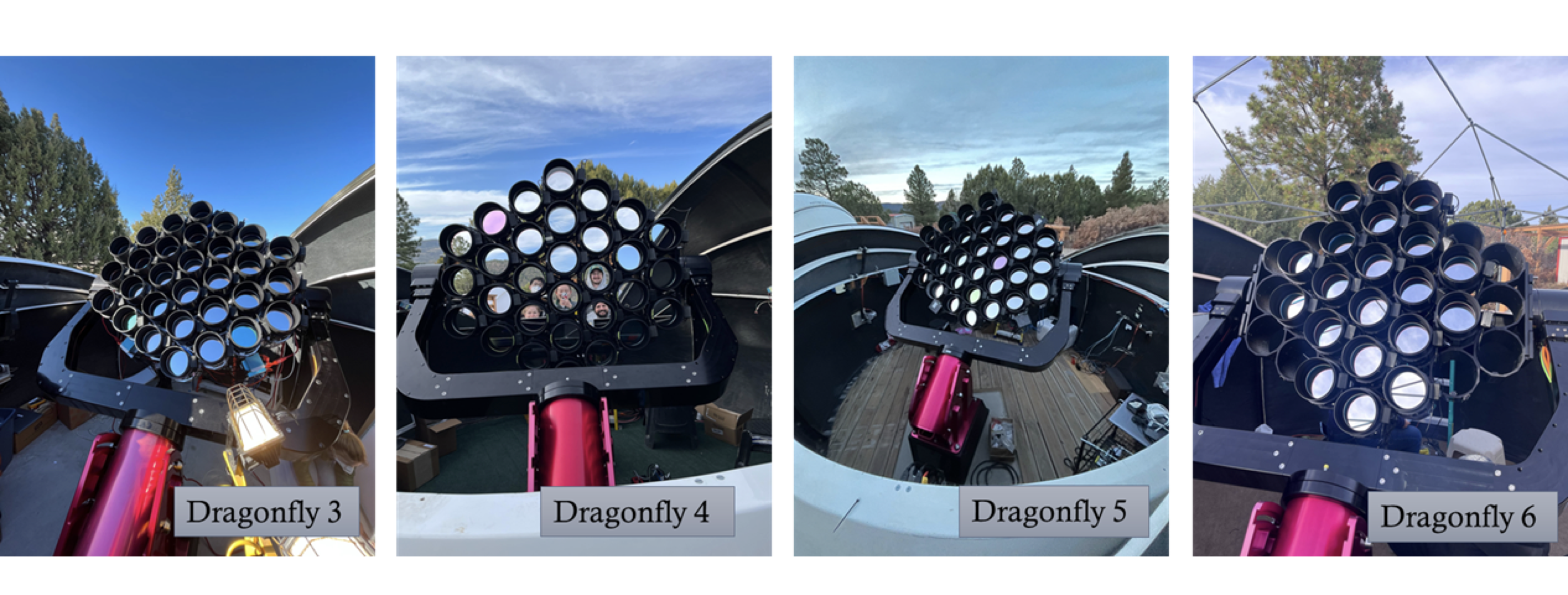}
    \caption{The four mounts comprising the DSLM during construction and commissioning (Mount 6 has several lenses not yet installed in this picture). Each Bisque Taurus Direct Drive Mount hosts a honeycomb of 30 lenses (with their associated cameras and peripherals).}
    \label{fig:mounts}
\end{figure}

The Dragonfly Spectral Line Mapper (DSLM) is a novel distributed-aperture telescope comprising 120 Canon telephoto lenses arranged into four mounts via a honeycomb array (Figure \ref{fig:mounts}). Equipped with ultra-narrowband filters (0.8 nm), the DSLM is designed to reach unprecedented depths in low surface brightness line emission features in the circumgalactic medium of nearby galaxies \cite{Lokhorst2019}. DSLM observes three key diagnostic optical emission lines: $H\alpha$, [NII], and [OIII], and is equipped with continuum filters along with other calibration filters to monitor the OH molecule driven skylines that are the major contaminant of the data. Designed to be operated in a remote, semi-to-fully autonomous manner, DSLM will conduct deep campaigns of nearby galaxies operating every clear night. Leveraging a large field of view and excellent sensitivity, the instrument, currently operating in New Mexico, is undertaking a survey of nearby galaxies to characterize their surrounding baryonic reservoirs (for detailed descriptions of the motivation and construction of DSLM, see Refs. \citenum{abraham2022,Lokhorst2020,Lokhorst2022SPIE,chen2022}). A pathfinder version of the instrument with three lenses and wider filters was also created, producing new science results from deep imaging of the M81 group \cite{pasha2021,Lokhorst2022shell}.

As described in Ref. \citenum{chen2022} and Chen 2023 (submitted), the fundamental unit of the DSLM consists of 1 Canon Telephoto lens, 1 Diffraction Limited SBIG Aluma CCD camera, 1 Diffraction Limited Starchaser unit, which connects the Aluma to the lens, passes through lens commands, and has a small pickoff mirror and detector for closed loop active optics, 1 Arduino device which manages communication and power to the lens, 1 FilterTilter, which uses a motor to tilt the ultra-narrowband filters in front of each lens, 1 Pegasus Pocket Powerbox Advanced Gen2, which supplies power to the rest of the components, and 1 Raspberry Pi, which controls the other hardware. This means that for the whole array, there are over 700 active hardware components with active connections and ongoing communication. In addition, the four mounts and their guiders, the mount PC computers, and the control computer, along with power controllers in each dome add additional network components. 

We adopt a hierarchical model for the software architecture managing this array of connections. Building from the bottom up, individual hardware elements are controlled from their unit's Raspberry Pi, which receives requests via an API over the local access network. Pis are then grouped by mount, enabling both array wide and mount specific commands to be sent. The key design philosophy is to create composable objects for each level of abstraction up from the hardware, each binding the objects one layer down as attributes and organizing them internally. This object-oriented approach provides the flexibility to simultaneously be able to send blanket commands to all units while also retaining the ability to drill down and assess the status of any individual hardware element on a particular unit at any time. 

There are five main layers in the DSLM software control system which we describe in this paper: 
\begin{enumerate}
    \item The hardware classes, which on each Pi handle communication with a given piece of hardware, 
    \item The web servers, which parse incoming API requests, call the hardware classes, then return responses, 
    \item The \texttt{DragonflyUnit} class, which binds hardware objects that mirror the API of the core hardware classes but call the relevant web server endpoints, 
    \item The \texttt{DragonflyManager} class, which binds wrapper classes which facilitate more complex commands to be sent across the array and handles unit management, and finally, 
    \item The autonomous observing software, which controls the instrument activity during the night.
\end{enumerate}

\begin{figure}
    \centering
    \includegraphics[width=\linewidth]{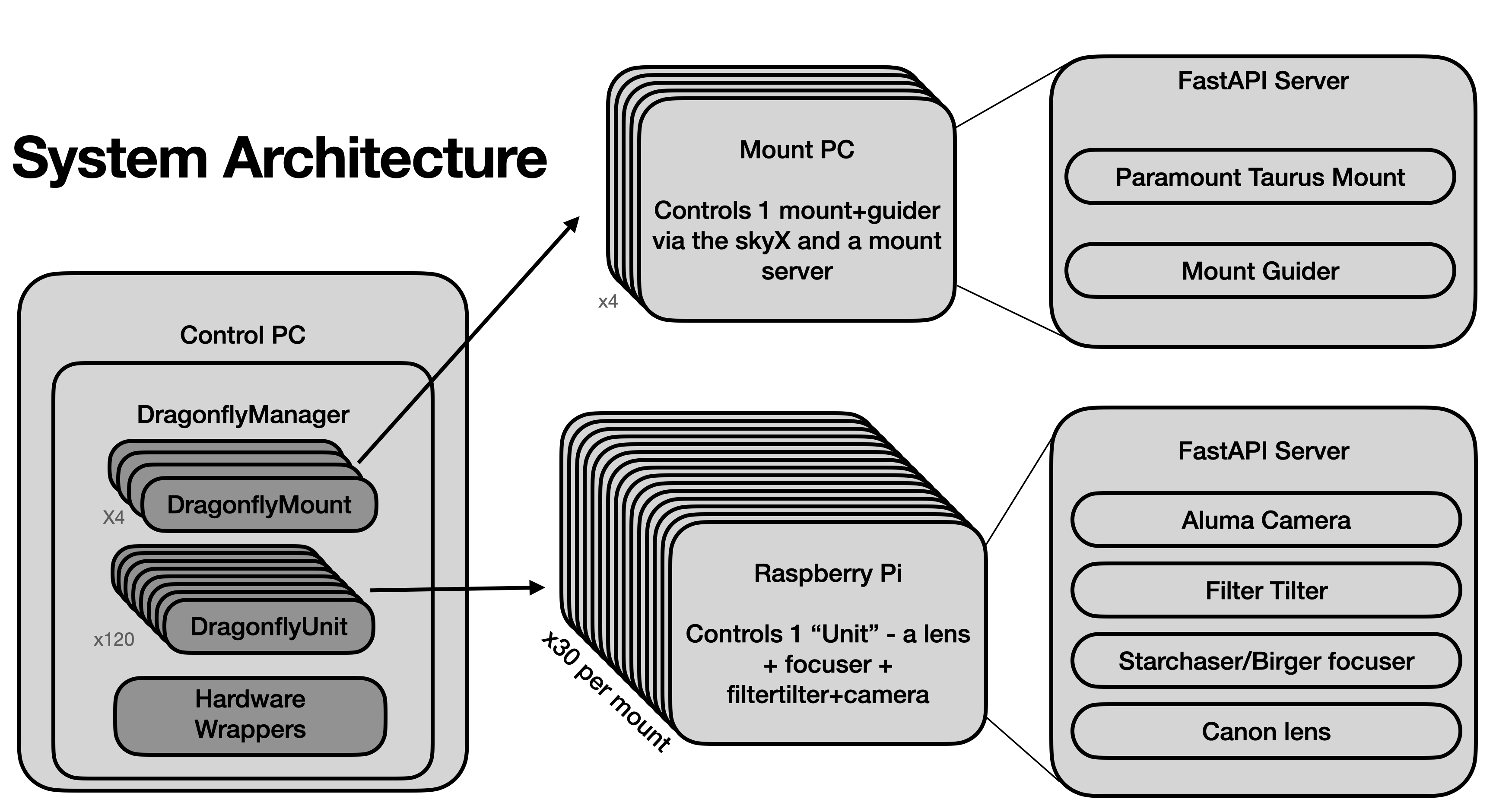}
    \caption{General layout of the computing setup for the 120 lens array. A single control computer runs the orchestrating observation code, via a manager Python class composed of individual unit classes, mount system classes, and convenience wrappers. These can be used to construct single or multi-unit commands, which are passed to \texttt{FastAPI} servers running within docker containers on either the 120 Raspberry Pis (for the units) or the Windows Subsystem Linux instances on four PCs (for the mounts). On the right, we show the primary hardware elements controlled by each, described in Section 3. }
    \label{fig:overview}
\end{figure}

An overview of these layers is presented in Figure \ref{fig:overview}, showing the composition of objects on the control computer side as well as the organization of servers across the devices.

The autonomous observing software makes use of the \texttt{DragonflyManager} object and its convenience wrapper methods to compactly make relevant requests (e.g., to slew, to guide, to expose, to tilt filters). The convenience wrapper methods construct their command arrays via calls to the \texttt{DragonflyUnit} objects, which can supply the formatting for requests relevant to each unit, by calling the appropriate hardware class contained within. The trade-off for this flexibility is added architectural complexity; we attempt to dampen this with the use of regular schemas and naming conventions, as we will discuss. 

In the following sections, we describe in more detail the various layers of the software stack and their purpose and motivation in the context of managing highly-distributed telescope systems.

\section{THE DRAGONFLY HARDWARE COMMUNICATION PROTOCOL}

At the very bottom of the software stack is the Dragonfly Communication Protocol, or DCP. This is a pure-Python schema (with the occasional use of libraries like cppyy to access C++ libraries) comprising a single object-oriented class for each piece of hardware, which handles low-level tasks like sending raw serial commands to the hardware and parsing the results. 

\subsection{Object-Oriented Hardware Control}

The \texttt{BaseHardware} Abstract Base Class Meta defines the minimum required structure for classes which are used for hardware interfacing. It defines several methods common to all DSLM hardware, including methods for
\begin{itemize}
    \item Connecting to the hardware (generally via a serial connection),
    \item Disconnecting from the hardware,
    \item Refreshing and returning the status of the hardware,
    \item Enabling polling on the hardware, and
    \item Disabling polling on the hardware.
\end{itemize}

For the DSLM system, the primary subclasses to this base class are \texttt{DLAPICamera}, which is an object responsible for interfacing with the Diffraction Limited SBIG Aluma and Starchaser cameras, \texttt{PegasusPowerbox}, which facilitates control of the Pegasus power controller which can be used to power cycle the other peripherals, \texttt{CanonEFLens}, which communicates with the Canon telephoto lenses, and \texttt{FilterTilter}, which communicates with the Dragonfly Filter Tilter motors. 

Each of these subclasses define additional methods relevant to their hardware control, e.g., exposing for cameras, tilting for filters, and setting focus positions for lenses. This system produces clean interface with the following API: 

\begin{verbatim}
import FilterTilter 
ft = FilterTilter(**kwargs)
response = ft.tilt_filters(10)
response = ft.get_tilt()
\end{verbatim}

\noindent Each class takes in several optional keywords for setting the port (e.g., /dev/ttyUSB1), but because USB ports are allocated dynamically, we generally use an internal method to ``find'' the appropriate port for each hardware element on startup. The classes for hardware in the array serve two primary purposes: to codify the communications with the hardware, and to store and furnish state information about the hardware. In general, these two tasks are related; executing hardware commands in general update the state of the hardware. 

Exception handling is a key element of the core hardware object methods, because ultimately these methods are being triggered within endpoints of a web server being routed with requests from a control computer, which is responsible for parsing hundreds of responses from variably successful units in a uniform way. To this end, all public methods (i.e., those which are exposed by the server endpoints, which we denote in our schema as all methods that do not start with an underscore) return a specific object: a \texttt{DragonflyResponse} with the following form: 
\begin{verbatim}
    DragonflyResponse:
        success: bool
        description: str
        payload: dict
        warnings: List[str]
\end{verbatim}

\noindent This schema is a flexible way to capture several key elements of a hardware request response: \texttt{success} to capture a top level assessment of whether the desired command was executed properly, a descriptive message about what was done, a payload dictionary to capture any return values (e.g., when requesting the CCD temperature or the filter tilt) or any errors, and a warning list of any non-critical warnings generated by the low level code during command execution. 

Ensuring that all public methods always return an object of this type ensures the orchestrating control software can uniformly parse all responses from all units, even in the case of errors. As such, the public methods, e.g., the filter tilters' \texttt{.get\_tilt()} are almost always wrappers around an internal method of the same name which actually communicates with the hardware. Exceptions raised by internal methods are then captured and added to the payload of the sanitized responses. 

Note that when using a framework like \texttt{FastAPI} (see Section 4), the types of quantities that can be added to, e.g., the payload dictionary must be in general JSON-serializable. Numpy arrays, for example, do not meet this criteria, nor do IEEE 754 floating point codewords such NaN and Inf. We implement a simple set of middleware functions to sanitize any response before sending it back over the network, turning arrays to lists, NaNs to None's, etc. 

These core hardware methods provide a flexible but intuitive user interface for directly controlling hardware elements on the array. It has proven invaluable for development, testing and debugging, while being easily composable into larger, more complex structures.

\section{Unit-Level and Mount-level REST API SERVERS}

As mentioned, some sort of framework is needed to facilitate the instantiation of the listed hardware objects on each unit of the array, and to subsequently call methods of those objects to carry out hardware tasks. Such commands need to be communicated over a local access network, which lends itself well to a web server of some kind to parse network requests and translate them into method calls. 

Because our core communication protocol is in Python, we select a Python web server framework, and after some consideration settled on \texttt{FastAPI} for its lightweight implementation, speed, and robust type validation features. We design our servers to have endpoints for all relevant hardware commands and to translate incoming requests directly into hardware class method calls. Upon startup, a single set of objects (one for each hardware element) are instantiated, which are then passed to the various routers for each hardware. Some hardware classes accept instances of a different hardware object, e.g., the camera object can add instances of the lenses and filter tilters, such that focus positions and tilts can be retrieved during exposures for header addition. 

The web server for DSLM is designed to be almost entirely ``logic-less''; that is, incoming HTTP requests over the network (generally containing either a JSON body or a URL-based request) are directly parsed, with keyword arguments being type verified before being passed directly to the relevant hardware object. A simple example is provided below. 

\begin{verbatim}
@router.post(filtertilter_schema.set)
def set(angle:float):
    return ft.set(angle)
\end{verbatim}

\noindent The route endpoint for setting the angle of the filter tilter expects an angle that can be parsed as a float. It then passes this argument directly to the \texttt{ft.set()} method, where \texttt{ft} is an instance of the \texttt{FilterTilter} hardware class. The server makes no computations, adjustments, or decisions regarding the way that the methods are called. This separates all logic about hardware interaction from the server; either the higher level functions on the control computer, or lower level sub-methods (e.g., internal class methods run within \texttt{ft.set()}) are responsible for any such decision making. This keeps the server as a clean translator which can be, for the most part, ignored from the perspective of debugging hardware commands.

We also note that all server endpoints \textit{directly return} the output of the Hardware Class method. \texttt{FastAPI} can send several forms of data over the network, from strings to float values to more complex JSON bodies. As described above, because all public-facing methods of the Hardware Classes return a \texttt{DragonflyResponse} object consisting of the attributes \texttt{success} (bool), \texttt{description} (str), \texttt{payload} (dict) and \texttt{warnings} (list), \texttt{FastAPI} parses this object (sanitizing values if needed) into a JSON format and returns it over the server. This ensures that when a command is sent to a server endpoint, the response will always be in the same format from all units --- either a timeout or error in the request itself, or a dictionary containing these exact keys. Exceptions raised during internal hardware methods are parsed by the public facing methods (generally, the raised exception is placed in the payload for examination). This style further adheres to the ``logic-less'' server principle, in which the endpoints neither make changes or decisions about the inputs, nor do they parse any outputs (besides the aforementioned middleware); they are simply a handshake between the local hardware and the requesting orchestrating computer. 

\subsection{Global Schema for Endpoints}

In the filter tilter example above, the \texttt{@router.post()} decorator is define via an imported module quantity: \texttt{filtertilter\_schema.set}. This syntax highlights a key organizational principle of the DSLM software stack: a single schema file for managing endpoint URLs. 

When adding features to the software stack, a method may be added to the relevant hardware class, then an endpoint constructed to serve that method. Rather than defining the URL endpoint directly in the server, we do so in a global schema file which may define it as, in this example, 
\begin{verbatim}
    router_prefix = "/filtertilter"
    set = "/set-tilt"
\end{verbatim}

\noindent Here, the router prefix indicates that all routes to this hardware go to a specific prefix, followed by the actual endpoint definition (we show set here, but all public endpoints are defined in this file). We then import the relevant string within the router file. In this case, because the router files define their prefix from this same file, the resulting endpoint will be ``/filtertilter/set-tilt''.

\begin{figure}
    \centering
    \includegraphics[width=\linewidth]{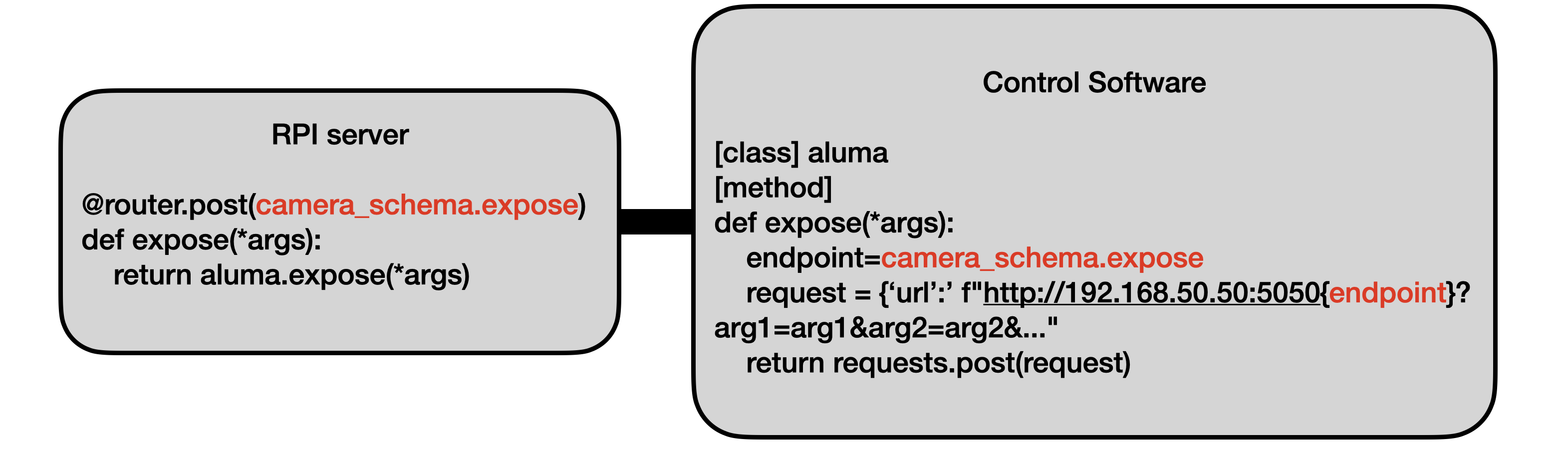}
    \caption{Use of a single ground-truth for endpoint definitions, defined both on the server and when constructing requests to those servers, ensures that route mismatches are never responsible for failed requests. Both the router post decorators (left) and the control software command generation code (right) import the same \texttt{camera\_schema} module, retrieving the same endpoints.}
    \label{fig:endpoints}
\end{figure}

This is useful because up a layer of abstraction, on the control computer, orchestrating commands need to be able to construct requests to the servers --- they need knowledge about the endpoint addresses. As we will see below, the constructor methods on the control PC will import the same endpoint strings, ensuring they are always consistent and correct, and that if changes to the endpoint syntax are desired, these can be implemented in one place and will be automatically adopted everywhere. This symmetry is shown in Figure \ref{fig:endpoints}.

\subsection{Server Hardware Routes}
The \texttt{main} module of each server has several endpoints defining a few simple methods, such as returning whether the server is active, or creating directories on the Raspberry Pi. The rest of the methods, which relate to the various pieces of hardware, are kept in individual router files which are imported to the main server file.

The router files trivially implement a series of endpoints named for their associated method call, e.g., 
\begin{verbatim}
@router.post(lens_schema.set_focus_position)
def set_focus_position(focus_value: float):
    fv_r = lens.set_focus_position(focus_value=focus_value)
    return fv_r


@router.post(lens_schema.set_is_x_position)
def set_is_x_position(is_x_position:float):
    isx = lens.set_is_x_position(is_x_position)
    return isx


@router.post(lens_schema.set_is_y_position)
def set_is_y_position(is_y_position:float):
    isy = lens.set_is_y_position(is_y_position)
    return isy


@router.post(lens_schema.set_zeropoint)
def set_zeropoint(zeropoint:float):
    zp = lens.set_zeropoint(zeropoint)
    return zp


@router.post(lens_schema.activate_image_stabilization)
def activate_image_stabilization():
    return lens.activate_image_stabilization()
\end{verbatim}

\noindent These endpoints (defining their URLs from the schema) simply pass arguments directly to the relevant method and return the result. The only source of additional complexity beyond this implementation within the server is for methods which require the passing of complex input data (e.g., facilitating the camera expose command to read in a header dictionary to add to the saved image headers). This is accomplished by taking advantage of the fact that \texttt{FastAPI} interfaces with \texttt{Pydantic} models. The following example illustrates this usage for the exposure case: 

\begin{verbatim}
from pydantic import BaseModel
class ExposureRequest(BaseModel):
    exptime: float
    n: int = 1
    imtype: str = "light"
    filename: Optional[str] = None
    checksum: bool = False
    debug: bool = False
    asynchronous: bool = False
    header_dict: Optional[Dict[str, Any]] = None
    savedir: Optional[str] = None
    analyze: Optional[bool] = True
    detection_sigma: Optional[float] = 9.0
    min_area: Optional[int] = 4
@router.post(camera_schema.expose)
async def expose(exposure_request: ExposureRequest):
    kwargs = exposure_request.model_dump()
    return aluma.wrap_exposure(**kwargs)
\end{verbatim}

\noindent In this (simplified) example, we define a \texttt{Pydantic} model (essentially a data class) which defines all the parameters to be passed along with their expected types. We then define the argument in the expose endpoint to take in an instance of this \texttt{ExposureRequest}. When defining inputs this way, the endpoint will expect the incoming HTTP request to contain a JSON body with this information, rather than encoding it into the URL.

\begin{figure}
    \centering
    \includegraphics[width=\linewidth]{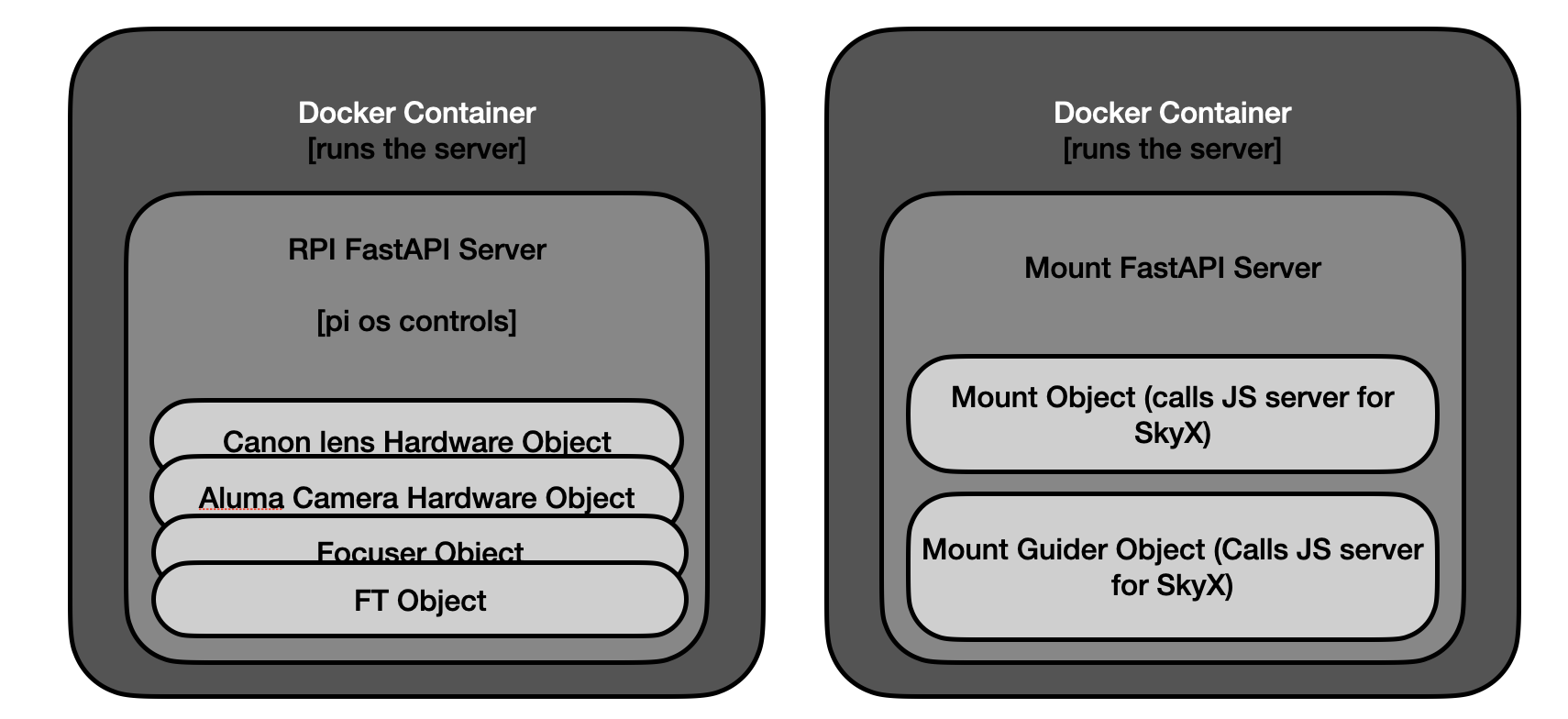}
    \caption{Overview of the software hierarchy on the Raspberry Pis (left) and Mount PCs (right). At the outermost layer is a docker container, within which all of the software dependencies and libraries are installed. When the docker container is spun up, it automatically runs the server startup command, which then starts the \texttt{FastAPI} server, which on startup creates the relevant hardware objects and uses their \texttt{connect()} methods to attempt to connect to them. Docker containers help ensure software consistency across the array, as well as allow for ease of container-level orchestration (e.g., restarting all unit servers).}
    \label{fig:docker}
\end{figure}

\section{DOCKER IMPLEMENTATION}
\subsection{Motivation}
Running software for large numbers of (nearly) identical systems, e.g., the units across the DSLM array, poses specific challenges. The computers controlling each unit must all have the requisite software dependencies needed to control the hardware, which must be installed and kept up to date. This step is challenging to keep consistent across many compute systems. Additionally, for reasons of security, the individual computers controlling the hardware across the distributed array do not have direct internet access, hampering the typical procedures for installing software dependencies from online repositories. 

The Docker containerization framework solves these challenges while providing additional orchestration benefits. By creating a Docker image (which is much like other operating system images), one packages everything needed to run certain code (from the operating system up), which can then be run as a form of virtual machine on any computer which has Docker installed. A Docker ``container'' refers to an instance of a Docker image running on a machine. Note that while this provides many isolation benefits, the underlying architecture of the machine is still being used by the container, so architecture based build parameters must be matched. In the case of DSLM, this means building for an ARM-64 architecture.

\subsection{Docker Development Cycle}
Our framework for building and deploying new code to the array can be summarized as follows.

\begin{enumerate}
    \item When new code which is part of the low level software control system is pushed to Github (or merged from a pull request), a Github Actions workflow is triggered. 
    \item This action workflow builds new versions of the Docker image for the codebase. This is a multi-stage build, which places the core operating system and library installations into one image which is then used as the base of a secondary image which involves only copying the control software over. This has the advantage that, most of the time, no changes are needed in the base image, resulting in build times of under two minutes. The final image is then pushed to DockerHub.
    \item From DockerHub, new images are pulled onto a central Raspberry Pi with ethernet access to the array, which has internet access. The release is then tagged and pushed to a local registry. 
    \item Finally, each unit pulls the updated image over the local network, then restarts their instance of the Docker containers (which utilizes the newly pulled Images).
\end{enumerate}

When a container is spun up on the Raspberry Pis (or Mount PCs), it automatically runs an initial \texttt{uvicorn} command to start the \texttt{FastAPI} servers. So in essence, restarting the containers amounts to restarting the servers. The organizational composition of the hardware objects, the servers, and the containers is summarized in Figure \ref{fig:docker}.

\subsection{Docker Container Orchestration}
Numerous increasingly complex tools are available to orchestrate many Docker containers at once. A simple, general purpose tool which can serve this purpose (along with the parallel execution of other system level tasks) is \texttt{Ansible}, which uses pure SSH connections to send simple commands (like starting, stopping, or restarting containers) to any number of units, specified in inventory files. For the purposes of DSLM, we have currently not seen the need to expand beyond \texttt{Ansible} for managing our containers across the array. Our primary needs regarding our containers is the ability to quickly update, restart, or bring up or down the containers on arbitrary subsets of units, and \texttt{Ansible} does this quite straightforwardly. This framework is best suited to the DSLM because while each container runs identical versions of the software and underlying libraries, each unit is not, in fact, identical. The primary functional difference between any two units is which filter is installed ($H\alpha$, [OIII], or one of the calibration filters), though there is also at current a mix of devices related to focusing the lenses. Requests are unit specific, so platforms which aim to create and manage containers as exact copies for the purposes of, e.g., load balancing incoming traffic (e.g., Kubernetes, Docker Swarm) are not quite quite appropriate. As a note, we do use the \texttt{Portainer} platform to provide additional dashboard / log-viewing capabilities with respect to our set of containers. If the size of the array grows, other options can be considered for orchestration purposes, but Docker as an industry-wide platform is well poised to either provide, or have ancillary tools which provide, the needed scalability. 

There are a few key considerations when attempting to utilize Docker containerization for the tasks outlined here. First, the Docker containers almost certainly need to be run in privileged mode, because (for good reason) containers are typically prevented from accessing hardware on their host machines (e.g., USB connected devices). This is obviously a requirement for controlling hardware from within a container. Additionally, to facilitate the saving of images to directories on the computer, or to load in relevant information, one must mount those files or directories during the docker-compose step to ensure code within the container can read and write to the appropriate locations. 

\section{OBSERVATORY MANAGEMENT SOFTWARE}

We approach the problem of orchestrating the behavior of hardware on units across the distributed telescope through a set of class compositions, which are treated as ``building blocks'' that allow fine tuned control to be expanded to generalized, all-unit commands.

\subsection{\texttt{DragonflyUnit} and \texttt{DragonflyMountSystem}: Object Oriented Hardware Control}

\texttt{DragonflyUnit} objects are defined by a name (e.g., Dragonfly301), which, from a configuration file, then sets several flags about the unit (e.g., whether it is a science unit or calibration unit, and more specifically, an $H\alpha$ unit or [OIII] unit, etc.). Bound to these objects are a set of methods which perfectly mirror those at the base hardware level in most cases. For example, if the Base Hardware \texttt{FilterTilter} class (an object which is used to run commands on the Pis) has a \texttt{set\_tilt()} method, then \texttt{Dragonfly301.filtertilter.set\_tilt()} exists at this high level. Each Dragonfly Unit has sub-objects which define the set of accessible hardware, and can use the \texttt{requests} library to directly issue commands to individual hardware elements on individual units in this manner. 

 We split the task of sending such commands into two: 
 \begin{enumerate}
     \item A private method named as \texttt{.\_methodname\_request()} is called, whose only responsibility is using known information about the unit, the method being called, and any needed parsing to construct a request to send. This is returned in the form of a dictionary with a URL key for the endpoint and either a URL style request, or a more complex request in which the `data' key holds a dictionary to be passed as a JSON body. 
     \item The public method (\texttt{.methodname()}) simply calls the private method to construct a command, then uses the \texttt{requests} package to send it, and returns the \texttt{DragonflyResponse} (a dictionary) to the user. 
 \end{enumerate}

 \noindent This splitting is necessary because when we introduce the \texttt{DragonflyManager}, the manager will need to generate commands for \textit{all} units at once, though each unit may have specific differences in the way such commands are constructed. For example, each unit has its own IP address, and method arguments may be parsed differently depending on the type of filter installed on the unit. The two step request generation $+$ sending framework means that the multi-command generation can take advantage of the same constructor that the individual unit level command does, and because they formulate requests in the same way, debugging is facilitated (the command can be tried in isolation on a single unit). 

 The \texttt{DragonflyMountSystem} objects operate similarly: they have the ability to generate requests using a method API that mimics that of the direct Hardware classes, and can post (or get) those requests to the matching mount's server. In the case of the mounts, the ``system'' comprises the Bisque Taurus mount (controlled by the the Windows software The SkyX) and the guider (also controlled by The SkyX). The mount servers themselves are simple wrappers which take input arguments and construct the Javascript snippets needed by The SkyX TCP server to send scripted commands. 

\begin{figure}
    \centering
    \includegraphics[width=\linewidth]{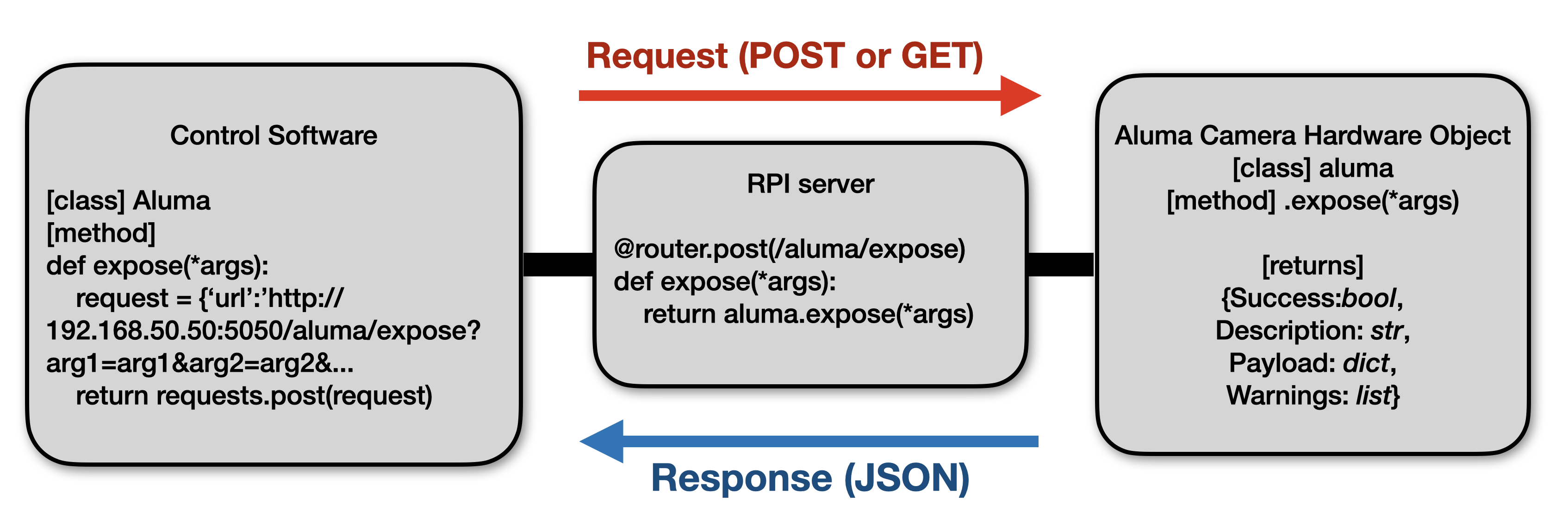}
    \caption{Overview of the chain of information passage by which a request is generated for a given unit (left), passed to the server via the local access network (center) and then passed to the (usually equivalently named) hardware class method, which then returns a DragonflyResponse that is parsed as a JSON body and passed back up the chain. In this case, the request is demonstrated as a URL request, but in practice, some commands, like \texttt{expose()}, utilize a JSON body to allow for the passage of more complex arguments (e.g., header dictionaries). Note additionally that we write out the endpoint here for clarity but in practice they are imported from our schema.}
    \label{fig:chain}
\end{figure}

\begin{figure}
    \centering
    \includegraphics[width=\linewidth]{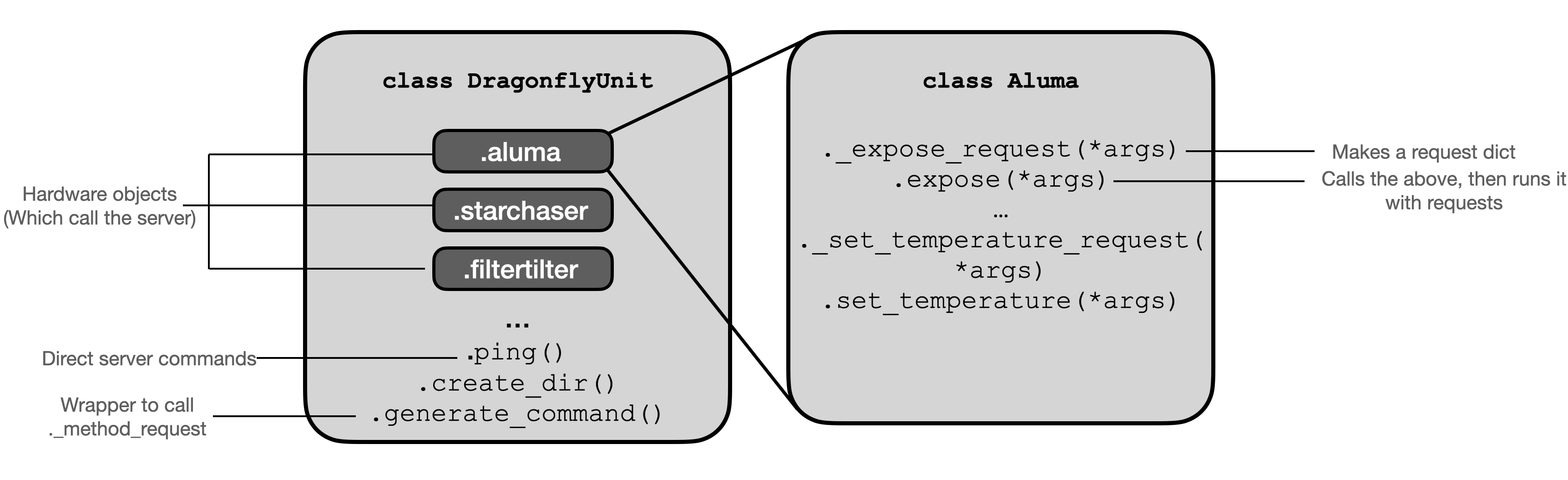}
    \caption{Examples of the methods present in the \texttt{DragonflyUnit} objects, which allow for individual units to be controlled from the top level, and are also simultaneously responsible for supplying the appropriate requests for a particular unit. In this hierarchical model, the same code used to generate a request for a command such as \texttt{Dragonfly301.aluma.expose()} is called when the manager is attempting to construct construct a full array of commands to all units. This is the reason for the existence of ``\_method\_request'' methods, which are called both by the higher level code and by the ``direct'' call to the method by the same name.}
    \label{fig:requests}
\end{figure}

An example of this mirrored API framework (and the server as the middleman) is presented in Figure \ref{fig:chain}, while Figure \ref{fig:requests} provides an example of how the \texttt{DragonflyUnit} itself is organized. This naming convention has the obvious benefit of making it clear what base level command is being run when a high level command is sent.
 
\subsection{The \texttt{DragonflyManager}: Command Aggregation and Asynchronous Request Management}

The \texttt{DragonflyManager} is a top level class which comprises all of the \texttt{DragonflyUnit} and DragonflyMountSystems active on the array (Figure \ref{fig:overview}). While they can be directly accessed as attributes (e.g., via a call to \texttt{manager.Dragonfly301.aluma.expose()}), the primary use of the manager is to construct commands to send to many units at once. It does so via a two step process: first, a method \texttt{retrieve\_subset()} is run which iterates through the extant flags for each \texttt{DragonflyUnit} (or mount system) and returns those which match input conditions. By default, this returns all units.  One can also specify `science' or `halpha' or `ha\_left' or `calibration' among several options, or a list of specific unit numbers to target, and one can provide a list of flags to select on any. The command also has the ability to skip units marked down by a global tracker, and to skip units manually passed in. 

Second, after composing a final set of units matching the input query, the \texttt{generate\_commands()} method uses the private request constructor method of each \texttt{DragonflyUnit} to construct a pandas DataFrame with each unit name and the request dictionary corresponding to that unit. Finally, a custom \texttt{RequestHandler} class takes this DataFrame, extracts the requests, and sends them using an asynchronous framework (in this case, \texttt{aiohttp}), allowing for all commands to be sent out to all units in parallel, while the handler then waits for all units to return (or time out) before supplying the user a DataFrame summarizing the results of the request. It should be noted that asynchronous API calls are not simply efficient but in fact a necessity for this array: with 120 lenses, even commands that take one second to run would take two minutes if the commands were sent to the array in serial. 

The default returned DataFrame is constructed from the dictionaries of actual responses, for which the keys are known because of the uniformity of the DragonflyResponse schema. However, for many commands, further parsing improves the usability of the returned values. Namely, when one or more quantities are requested from the array, they are stored in the `Payload' dictionary within the Response dictionary. This results by default parsing in a DataFrame with a column `Payload' which has, as each entry, a dictionary of desired information. 

To flexibly and reliably convert these into more useful data structures, we employ a YAML-based schema. When a command key is passed to the \texttt{RequestHandler}, the response DataFrame is additionally parsed to set desired payload keys as full columns of the DataFrame. 

As a concrete example, for a call to retrieve the temperatures of the Aluma CCDs, this converts a DataFrame of the form 

\begin{verbatim}
Unit          Success Description Warnings Payload
Dragonfly301    True         None     None {'temperature':...}
Dragonfly302    True         None     None {'temperature':...}
...
\end{verbatim}

into 

\begin{verbatim}
Unit          Success  Warnings Temperature
Dragonfly301    True       None       -13.5
Dragonfly302    True       None       -11.2
...
\end{verbatim}

\noindent The response schema YAML file simply defines which columns are desired in the final DataFrame and which payload keys to convert into new columns, e.g., 

\begin{verbatim}
get_temps:
  required_columns:
     Unit: str
     Success: bool
     Payload: dict
     Warnings: list
  payload_keys: 
     temperature: numeric 
\end{verbatim}

\noindent The simple definitions allow for some customization to be applied to the returned, aggregated DataFrames while retaining the same parsing code. Note that we also carry out type validation using this schema, so when units return incorrect values (or no values), this can be sanitized in order to not raise exceptions. This parsing makes the analysis of the results of a command easier to carry out; in the case above, a decision might be made about marking units with aberrant sensor temperatures as ``down'' and disabling their cooling. It is also more user-friendly during testing to create columns of the desired response values (printouts of DataFrames often truncate the information subsumed in dictionaries within columns). It is also useful for the log files generated during automated observing procedures. For example, the focus of each unit is often assessed, and the parsed DataFrame can be written easily to the log for inspection: 
\begin{verbatim}
    Unit         ... Success skymedian         fom    fwhm  fwhm_rms  nobj  FocusGoodness
0   Dragonfly301     True        1999.0   25.683975   1.791     0.466    46           GOOD
1   Dragonfly302     True        1997.0   32.472325   1.355     0.208    44          GREAT
2   Dragonfly303     True        2099.0  277.436946   1.467     0.331   407          GREAT
3   Dragonfly305     True        1997.0    1.831502   2.730     0.139     5       MARGINAL
4   Dragonfly306     True        2004.0   26.136364   1.760     0.468    46           GOOD
5   Dragonfly307     True        1975.0  -99.000000 -99.000   -99.000   -99           LOST
6   Dragonfly309     True        2002.0   25.390625   1.536     0.257    39          GREAT
7   Dragonfly310     True        2083.0  272.388060   1.876     0.442   511           GOOD
... 
\end{verbatim}

\subsection{Wrappers}

The final layer of abstraction in the DSLM codebase is the so-called ``hardware wrappers'' which are ultimately responsible for creating user friendly multi-unit calls on behalf of the observer. Up to this point, we have seen how an individual command can be issued to the hardware on an individual unit via the \texttt{DragonflyUnit} call (whether bound to the manager or not). 

To facilitate command distribution across the array, we define a final set of classes which all accept a \texttt{DragonflyManager} object as an input. These classes are named as the plurals of the hardware types (\texttt{Lenses}, \texttt{Alumas}, \texttt{Filtertilters}, etc.), as they are designed to be called when sending commands to multiple units at once. We then instantiate these objects as attributes of the \texttt{DragonflyManager}, meaning the manager injects itself as a dependency to the hardware wrappers. 

In the simplest cases, the hardware wrapper for a given single command involves only a few steps:
\begin{itemize}
    \item assessing which units to send the command to, via the ``which'' parameter passed to all method calls, along with any units to skip, and whether to only select units marked up,
    \item using the \texttt{manager.generate\_commands()} method to construct a DataFrame with those parameters along with any input parameters to the underlying function, and
    \item if a dry run is indicated, returning this DataFrame, or, if not, using the \texttt{RequestHandler} to execute the command and return the parsed response (either the simple parsing or more complex parsing as described above, based on whether a command key is supplied). 
\end{itemize}

\noindent An example of a simple command is a convenience function, \texttt{simple\_expose\_cameras}, which in a logic-less manner sends a single exposure time to all selected units. While selection criteria can affect which units are included in the command, at a fundamental level the constructed DataFrame is homogeneous. 

In contrast, many of the commands we need to construct are heterogeneous with regard to different units on the array. The convenience wrapper \texttt{set\_tilts(halpha\_tilt,oiii\_tilt)}, as the name suggests, allows the $H\alpha$ tuned filters to tilt to a different value than the [OIII] filter units. Within the method, this is handled by creating two separate DataFrames using the method above, one with ``which'' set to the $H\alpha$ flag, the other to [OIII]. The two DataFrames are then concatenated and sent out to the units. Beyond even this, some convenience wrappers even send different commands altogether to different units. This is generally discouraged, but is used on occasion: for example, below we will discuss an efficiency technique by which science exposure commands will take science exposures for in focus units, but conduct focus runs for out of focus units at the same time. The complexities of constructing such commands to send is handled within these wrappers. 

A final, common complexity addition to the basic wrappers is that of validation. For example, the convenience wrapper for setting tilts not only sends different tilt values to different units, but if desired, will then execute a check of the obtained tilts after the move, assess which units may have tilts out of a desired tolerance of the input tilt goal, and try to tilt those filters again. This type of set-check-retry pattern is common throughout the instrument, and vital for enabling autonomous observations. The wrapper methods provide the needed flexibility to construct both the complex command sets discussed above as well as to handle multiple rounds of commands as part of more complex procedures.

\subsection{Autonomous Observing}

DSLM takes advantage of the machinery built up from the core hardware classes, to the safety-oriented server request formats, to the \texttt{DragonflyUnit}, and ultimately to the \texttt{DragonflyManager} object to carry out autonomous observations every clear night. Here we describe the state machine setup of the instrument and how it facilitates automated observations. The goal of all code which is present across the categories listed below is to handle the logic of observing, without having to deal with the complexities of interacting with the hardware. The hardware wrappers discussed above facilitate this separation; in general, every ``state'' of the instrument as described below will prescribe a certain action, generally codified within one or several calls to a \texttt{manager.wrapper()} commands. Any needed complex handling is done within the hardware wrapper methods themselves, such that bugs during autonomous observations can be generally defined entirely as bugs in the state machine logic (what the instrument is \textit{attempting} to do) or communication/hardware logic (what the instrument is \textit{actually} doing in response to a given request). 

For the DSLM, we implement a ``state machine'' framework for managing the observatory. A state machine is a programming paradigm in which at any given time, the instrument can only be in a single, generalized state, and predefined triggers can cause that state to change. At the expense of each state having to be non-sequential (i.e., every time the instrument enters a state, it is the same as any other time), this framework enables a simplified algorithm for determining what the instrument should be doing at any given moment which does not depend on what it was doing in the past. Because the assessment of what the instrument should be doing depends on many factors, including the time of night, moon status, dome status, focus status, etc., we elect to use a loop of live assessment based on current conditions, which trigger new states which execute a given command or set of commands, then return to the assessment state. 

As we will discuss below, our observing sequence during the night is not \textit{entirely} free of sequence-based criteria. For example, as the instrument cycles through states of exposing on the target, it is expected to ``dither'' by a certain amount for data reduction purposes. Additionally, if we re-assess what target to observe after every exposure, we do not need to re-tilt the filters to a target's tilt if the target has not changed. We generally handle these slightly asymmetries via simple attribute checks at the front of each state method; e.g., we track the current target and can check if the tilt command actually needs to be run. 

In the following sections, we describe the pre-observing instrument health checks and the three separable state machine implementations for the startup phase, observing, and shutdown phase of each night.

\subsubsection{Pre-Observing}
In the late afternoon, before observing commences, a set of health checks is performed on the instrument, ensuring the hardware is connected, accessible, and responsive. Units not meeting certain criteria are marked down at this stage, and a report is generated of which units were marked down and for which test. When possible, remote observers check this report before sunset and trouble shoot any easily addressed issues, many of which can be resolved with a either a container/server restart or a power cycle to that unit. 

\subsubsection{The \texttt{DatetimeManager}, Target Class, and Celestial Almanac}

Three primary inputs are required for setting up nightly observing. A target list, consisting of \texttt{Target} objects that contain information about the targets (name, coordinates, start and end of desired observing windows), an Almanac, which computes the various astronomical times relevant for observing along with moon rise and moon set, and the \texttt{DatetimeManager}, which computes the current time, and facilitates computations surrounding time deltas. 

Critically, the Almanac class uses the \texttt{DatetimeManager} to compute all times or checks, and the \texttt{DatatimeManager} has a ``simulation'' mode in which any time and date can be simulated, and time can be run at arbitrary speeds. This ability allows for nights of observing to be realistically simulated, ensuring the procedures chosen by the autonomous software match expectations.

\subsubsection{The \texttt{StartupManager}}

The \texttt{StartupManager} is a state machine which handles the state of the facility from pre-sunset, when the nightly script is started, to the point at which observations of astronomical targets begin. The states covered by this are: 
\begin{itemize}
    \item  INITIALIZING,
    \item    WAITING\_FOR\_DOME\_OPEN,
    \item    WAITING\_FOR\_SUNSET,
    \item    ASSESSING\_CAMERAS,
     \item   ASSESSING\_NEEDED\_ACTION,
    \item    WAITING\_FOR\_TWILIGHT\_FLAT\_LEVELS,
    \item    TAKING\_TWILIGHT\_FLATS,
    \item    WAITING\_FOR\_12\_DEGREE,
    \item    SCIENCE\_HANDOVER, and
    \item    SHUTDOWN. 
\end{itemize}

\noindent The final two states actually initialize and run other state machines, which handle observing and shutdown procedures (including darks), respectively. When the conditions have not triggered one of the final two states, the instrument will move between states following set transitions. Initialization is always the first step, which will trigger either a hold for sunset, a hold for dome open, a shutdown, or an assessment of needed action. During initialization, several last checks are made, including that lenses are initialized and cameras are cooled, if needed. 

The hold states (e.g., hold for sunset and hold for dome open) run in loops which assess the requisite status every second. The dome state also contains checks at the end of each iteration for whether a bail out time has been reached (generally an hour before morning twilight) at which point the instrument initiates shutdown. 

Whenever the instrument exits a hold state, it returns to assess the needed action state. This method checks that initialization tasks are complete before setting the next state based on the time: for example, before 12-degree twilight, it will attempt the auto-flat routine, if requested, otherwise it holds for 12 degree twilight. At this point, if the dome is open, it initiates the handover to the Observing Manager. 

\subsubsection{The \texttt{ObservingManager}}

The \texttt{ObservingManager} class operates as a state machine whose valid states cover the duration of science observations during the night. Its valid states are 
\begin{itemize}
    \item  WAITING\_FOR\_DOME\_OPEN,
    \item     ASSESSING\_NEEDED\_ACTION,
     \item    WAITING\_FOR\_18\_DEG,
     \item    HOLDING\_FOR\_TARGET\_WINDOW,
     \item    TWILIGHT\_FOCUS\_ROUTINE,
     \item    SETTING\_UP\_FOR\_TARGET,
     \item    FOCUSING,
     \item    ASSESSING\_FOCUS,
     \item    SLEWING\_TO\_TARGET,
      \item   OBSERVING\_SEQUENCE, and
          \item     SHUTDOWN,
\end{itemize}

\noindent Like the startup manager, the hold states make periodic checks and have bailout times. The core method/state is again the assessing of the needed action. At any state change, the dome status is checked, as well as whether it is close to morning twilight and whether the observing window for a target has ended (at which point a new target is selected or a hold for next target is performed). 

In general, if the domes are open at 12-degree, the instrument will conduct focus runs until 18-degree twilight to get as many units in focus as possible. These dynamic focus runs establish the focus goodness of each unit with a test exposure, then choose coarse to fine focus windows based on that assessment. Once 18-degree twilight is reached, the code moves to the setting up for target state, in which the filters are tilted to the current target's tilt values. This state checks if the incoming target is the same as the current target, so tilts are only set when new targets are selected (due to the current time being in a different target's defined observing window). 

Focus assessment is carried out by taking a short exposure and extracting the sources in the image, which is carried out in threads on each Raspberry Pi at the end of an exposure. These values are then added to the camera state, and can be retrieved. Figure of Merit values are computed as the number of detected objects divided by the median full-width-half-maximum (FWHM) of detected sources in the image. Focus runs choose the focus position with the greatest value of this quantity; during assessment, only the FWHM value is used. We assign lenses a focus-goodness value of either great, good, marginal, poor, or lost based on FWHM thresholds. Lost focus indicates that no sources are found or that the number of sources found are under a minimum threshold value.

After slewing to a target, the focus is assessed. If more than a fractional threshold (currently 0.68) of lenses have good or better focus, then an exposure commences. However, because our science exposures are 900 s and our focus runs are $\sim$400 s, we do not expose on target with any units with marginal or worse focus; instead, these units conduct a focus run while the good/great focus lenses obtain science data. If instead less than the threshold of lenses are good or great, all lenses undergo the aforementioned dynamic focus runs. Thus, units already in good focus only explore small focus ranges in their vicinity, while out of focus units explore a much wider range. Further planned improvements to this particular algorithm involve having each unit choose the minimum of a parabolic fit to the focus values; this can be somewhat noisy and unstable in practice, however. 

A single observing sequence, which can be triggered by an OK recommendation from the focus assessment, consists of dithering on the target, starting the guider, exposing for the science duration (set in each \texttt{Target} object), then disabling the guider. The nature of the state machine means that the state then returns to ASSESSING\_NEEDED\_ACTION, at which point assessments are made about the current time in relation to evening or morning twilight or moon rise/sets, and to whether the current target should be updated. If the current target remains the same -- requiring that the observing window has not been left -- then the next state (setup on target) does nothing, and the following state (slewing to target) is executed. This is desirable, as the dithering is then handled always from the target center. After this, the focus is assessed, and so on. 

The carrying out of observations in this state machine context has several benefits. First, it is robust against the key time-related interruption faced by the instrument: dome closure. One can in principle add checks verbosely to catch \textit{when} the dome has closed (due to clouds or other inclement weather) to sequential code. But when the dome then reopens, it is very challenging to properly come out of that state and resume observations without significant decision trees. Here, the dome closing triggers a specific state which then triggers a clean instrument assessment state open reopening. If the target changed while the dome was closed, or twilight is approaching, these are smoothly caught. Second, the state machine allows code for defining how the instrument behaves to be siloed into separable functions; debugging the behavior of a particular state thus happens only in one place. 

The trade-off is that the states must be independent yet able to run in sequence over and over. In the case presented above, a specific check was thus required to prevent unnecessary tilting of the filters every time the exposing sequence proceeded on the same target. Conversely, a modulo-type operator was required to cycle appropriately through dither values in an otherwise independent sequence. In principle, such a behavior could be well-captured by a hierarchical state machine (in some sense, the division of states between a startup, observing, and shutdown manager represent a hard-coded hierarchical state machine), but in practice we have found that for our system the amount of additional boilerplate checks needed to skip certain steps given the instrument current state do not ultimately detract from the benefits of the state machine framework. 

Several different conditions could trigger the shutdown procedure (its own state machine) during the night, including: 
\begin{itemize}
    \item the dome is closed, and it is $<$1 hour from morning twilight. This condition captures the fact that restarting observations and getting into focus would not leave enough time for reasonable science observations,
    \item we finish a focus run or exposure and discover it is past morning 18-degree twilight, or
    \item we finish a focus run or exposure and discover we are past the end of the latest target window amongst our targets; there is nothing left to observe. 
\end{itemize}

\noindent In any of these cases, the \texttt{ShutdownManager} is initialized, passed relevant information, and executed. 

\subsubsection{The \texttt{ShutdownManager} }
The final state machine class handles shutdown / end of night procedures for the array. This includes 
\begin{itemize}
    \item PARKING\_MOUNTS,
    \item TAKING\_DARKS,
    \item DEREGULATING\_CAMERAS,
    \item CREATING\_NIGHTLY\_ASSESSMENT,
    \item SYNCING\_TO\_AWS, and
    \item DONE.
\end{itemize}

\noindent This set of states is relatively straightforward, and states always progress sequentially. After parking the mount, the array takes several darks matching the exposure times of any lights taken, either science or flats. Then the cameras are commanded to cease active cooling, and all the data gathered for the night is assessed (using focus information from the headers), going into a report issued by the instrument, including the targets observed, effective, in-focus frames/hours on each. Finally, the data and headers are synced to Amazon Cloud Services, which is the current database and storage solution used by the DSLM. 

\subsection{System Safety}

When working with multiple arrays of expensive hardware, safety is a critical concern. Many of the software schemas presented throughout this paper were chosen specifically to minimize the occurrence of uncaught exceptions along the chain of command. However, as not every exception can be caught \textit{in situ}, we implement a ``safe observe'' mode in which the full nightly script is run within a try-except block. Any exceptions raised that are not already caught or accounted for trigger the mounts to park and the cameras to cease active cooling, preventing the mounts from tracking indefinitely. 

More insidious are undesired behaviors which do not trigger exceptions. An important consideration is to not track the mounts over toward the floor; this can happen if checks about leaving an object's observing window are not frequent enough. A common issue in older iterations of this observing software was that focus runs occurred \textit{in situ} within the code, just before exposures, and could retry several times before giving up. If such a set of runs started just before twilight or the end of an object window, observations could continue in the undesired state for as long as an hour, In this new state machine framework, each single unit of action is a state, and all the conditions we need frequently checked are thus carried out both between every state, as well as \textit{during} any states which are of the holding variety. 

Additional concerns have implementations within the code base that are not discussed in detail here; for example, if the control PC were to crash, the mount servers (on the independent mount PCs) also have time-based hard stops that will tell the mount to stop tracking if it is sunrise. 

Ultimately, a certain amount of risk is presupposed whenever allowing a telescope to observe autonomously. We benefit from having observatory staff on hand at the facility to check in if things go wrong and to spot check the instrument health. Efforts are underway to actively monitor and alert on the logs being generated by all the hardware and software throughout the array in an automated fashion.

\section{Summary and Future Directions}

In this paper we have presented an overview of the various software architectural principles employed to effectively manage a highly distributed array telescope. We present a framework that leverages Pythonic object-oriented programming principles, including inheritance and composition, to organize each level of the software stack, enabling the construction of complex, heterogeneous commands to the full array, or individual commands to individual hardware elements on a single unit, all with equal ease. We additionally discussed several technologies which have facilitated uniform running of and access to our source code, namely, Docker and Ansible. Finally, we presented the state machine approach being used to drive autonomous observations. 

There are several ways in which the software stack could be extended in the future. The ability for the hardware to poll continuously provides the opportunity for live-updating dashboards of the entire array's status to be created, which we are in the process of integrating. Similarly, there is an initiative to develop from semi-autonomous observing (in which the targets for a given night, and their observing windows, are specified by an ``observer'') to fully autonomous surveying, in which a given set of survey targets are dynamically chosen for observing given various criteria. Ultimately, our goal has been to create a software stack that remains modular and separable at many levels, creating maximal flexibility for future development.

\acknowledgments 
 
The authors thank Chloe Neufeld for useful discussions on this work. The authors acknowledge the staff of New Mexico Skies, all of whom are credited with keeping the instrument healthy and able to observe. 

\bibliography{report} 

\begin{thebibliography}{10}

\bibitem{Abraham:2014}
{Abraham}, R.~G. and {van Dokkum}, P.~G., ``{Ultra-Low Surface Brightness Imaging with the Dragonfly Telephoto Array},'' {\em Proceedings of the Astronomical Society of the Pacific}~{\bf 126},  55 (Jan. 2014).

\bibitem{vanDokkum:2018}
{van Dokkum}, P., {Danieli}, S., {Cohen}, Y., {Merritt}, A., {Romanowsky}, A.~J., {Abraham}, R., {Brodie}, J., {Conroy}, C., {Lokhorst}, D., {Mowla}, L., {O'Sullivan}, E., and {Zhang}, J., ``{A galaxy lacking dark matter},'' {\em Nature}~{\bf 555},  629--632 (Mar. 2018).

\bibitem{Lanzetta2023a}
Lanzetta, K.~M., Gromoll, S., Shara, M.~M., Berg, S., Valls-Gabaud, D., Walter, F.~M., and Webb, J.~K., ``Introducing the condor array telescope. 1. motivation, configuration, and performance,'' {\em Publications of the Astronomical Society of the Pacific}~{\bf 135} (Feb 2023).

\bibitem{Law:2022}
{Law}, N.~M., {Corbett}, H., {Galliher}, N.~W., {Gonzalez}, R., {Vasquez}, A., {Walters}, G., {Machia}, L., {Ratzloff}, J., {Ackley}, K., {Bizon}, C., {Clemens}, C., {Cox}, S., {Eikenberry}, S., {Howard}, W.~S., {Glazier}, A., {Mann}, A.~W., {Quimby}, R., {Reichart}, D., and {Trilling}, D., ``{Low-cost Access to the Deep, High-cadence Sky: the Argus Optical Array},'' {\em Proceedings of the Astronomical Society of the Pacific}~{\bf 134},  035003 (Mar. 2022).

\bibitem{chen2022}
{Chen}, S., {Lokhorst}, D.~M., {Shen}, J., {Pasha}, I., {Malakhov}, E.~I., {Abraham}, R., and {van Dokkum}, P., ``{The Dragonfly Spectral Line Mapper: design and first light},'' in [{\em Ground-based and Airborne Telescopes IX}{\nolinebreak\hspace{0.1em}]},  {Marshall}, H.~K., {Spyromilio}, J., and {Usuda}, T., eds., {\em Society of Photo-Optical Instrumentation Engineers (SPIE) Conference Series} {\bf 12182},  121824E (Aug. 2022).

\bibitem{merkel2014docker}
Merkel, D., ``Docker: lightweight linux containers for consistent development and deployment,'' {\em Linux journal}~{\bf 2014}(239),  2 (2014).

\bibitem{Lokhorst2019}
{Lokhorst}, D., {Abraham}, R., {van Dokkum}, P., {Wijers}, N., and {Schaye}, J., ``{On the Detectability of Visible-wavelength Line Emission from the Local Circumgalactic and Intergalactic Medium},'' {\em Astrophysical Journal}~{\bf 877},  4 (May 2019).

\bibitem{abraham2022}
{Abraham}, R.~G., {van Dokkum}, P., {Lokhorst}, D., {Chen}, S., {Liu}, Q., {Rice}, M.~L., and {Rice}, E.~L., ``{Distributed aperture telescopes and the Dragonfly Telephoto Array},'' in [{\em Ground-based and Airborne Telescopes IX}{\nolinebreak\hspace{0.1em}]},  {Marshall}, H.~K., {Spyromilio}, J., and {Usuda}, T., eds., {\em Society of Photo-Optical Instrumentation Engineers (SPIE) Conference Series} {\bf 12182},  121821W (Aug. 2022).

\bibitem{Lokhorst2020}
{Lokhorst}, D.~M., {Abraham}, R.~G., {van Dokkum}, P., and {Chen}, S., ``{Wide-field ultra-narrow-bandpass imaging with the Dragonfly Telephoto Array},'' in [{\em Ground-based and Airborne Telescopes VIII}{\nolinebreak\hspace{0.1em}]},  {Marshall}, H.~K., {Spyromilio}, J., and {Usuda}, T., eds., {\em Society of Photo-Optical Instrumentation Engineers (SPIE) Conference Series} {\bf 11445},  1144527 (Dec. 2020).

\bibitem{Lokhorst2022SPIE}
{Lokhorst}, D.~M., {Chen}, S., {Pasha}, I., {Shen}, J., {Malakhov}, E.~I., {Abraham}, R.~G., and {van Dokkum}, P., ``{The pathfinder Dragonfly Spectral Line Mapper: pushing the limits for ultra-low surface brightness spectroscopy},'' in [{\em Ground-based and Airborne Telescopes IX}{\nolinebreak\hspace{0.1em}]},  {Marshall}, H.~K., {Spyromilio}, J., and {Usuda}, T., eds., {\em Society of Photo-Optical Instrumentation Engineers (SPIE) Conference Series} {\bf 12182},  121821T (Aug. 2022).

\bibitem{pasha2021}
{Pasha}, I., {Lokhorst}, D., {van Dokkum}, P.~G., {Chen}, S., {Abraham}, R., {Greco}, J., {Danieli}, S., {Miller}, T., {Lippitt}, E., {Polzin}, A., {Shen}, Z., {Keim}, M.~A., {Liu}, Q., {Merritt}, A., and {Zhang}, J., ``{A Nascent Tidal Dwarf Galaxy Forming within the Northern H I Streamer of M82},'' {\em Astrophysical Journal Letters}~{\bf 923},  L21 (Dec. 2021).

\bibitem{Lokhorst2022shell}
{Lokhorst}, D., {Abraham}, R., {Pasha}, I., {van Dokkum}, P., {Chen}, S., {Miller}, T., {Danieli}, S., {Greco}, J., {Zhang}, J., {Merritt}, A., and {Conroy}, C., ``{A Giant Shell of Ionized Gas Discovered near M82 with the Dragonfly Spectral Line Mapper Pathfinder},'' {\em Astrophysical Journal}~{\bf 927},  136 (Mar. 2022).

\end{thebibliography}
\bibliographystyle{spiebib} 

\end{document}